\documentclass[aps,prb,preprint,longbibliography,amssymb]{revtex4-1}
\usepackage[dvips]{harpoon}
\usepackage{graphicx}
\DeclareRobustCommand{\myorh}{\overrightharp} 
\usepackage{bm}
\begin{document}

\title{Orbits of the Kepler problem via polar reciprocals}

\author{E. D. Davis}
\email[Electronic mail: ]{davis@kuc01.kuniv.edu.kw}

\affiliation{Department of Physics, Faculty of Science, Kuwait University, P.O. Box 5965, 13060 Safat, Kuwait}

\date{\today}

\begin{abstract}%
It is argued that, for motion in a central force field, polar reciprocals of trajectories are an elegant
alternative to hodographs. 
The principal advantage of polar reciprocals is that the transformation from a 
trajectory to its polar reciprocal is its own inverse. The form of polar reciprocals $k_*$ of Kepler 
problem orbits is established, and then the orbits $k$ themselves are shown to be conic sections
using the fact that $k$ is the polar reciprocal of $k_*$.
A geometrical construction is presented for the orbits of the Kepler problem starting 
from their polar reciprocals. No obscure knowledge of conics is required to demonstrate the validity of the 
method. Unlike a graphical procedure suggested by Feynman (and amended by Derbes), the algorithm based on 
polar reciprocals works without alteration for all three kinds of trajectories in the Kepler problem 
(elliptical, parabolic, and hyperbolic). 
\end{abstract}

\maketitle

Approximately, a hodograph is a plot of velocities along a trajectory; more precisely, it is the locus of
the tips of the velocity vectors after they have been parallelly transported until their tails are at the 
origin (in velocity space). There have been many articles on the pedagogic virtues of 
hodographs.\cite{Abelson,Gonzalez,Apostolatos,Butikov,Mungan} \emph{Feynman's Lost Lecture}\cite{FLL} 
contains, amongst other things, a recipe for drawing the elliptical path of a planet given its hodograph. 
The procedure, as reproduced in Ref.~\onlinecite{FLL}, has some shortcomings,\cite{Griffiths,Weinstock} but, 
fortunately, these have been more than satisfactorily rectified by Derbes.\cite{Derbes} 
Derbes notes that Feynman's scheme also works for hyperbolic orbits and he is able to devise another 
construction for the exceptional case of parabolic orbits.
A completely different way of tracing all three kinds of orbits with the help of hodographs 
has been successfully developed by Salas-Brito and co-workers in a series of publications culminating in 
Ref.~\onlinecite{SalasBrito}. 

The various geometrical methods of the previous paragraph are easily implemented, but, to prove their
validity, a student would have to be acquainted with many properties of conics which no longer form 
part of most school curricula. 
Indeed, at least one reader of Derbes' article feels that his exposure
to the hodograph left him ``disappointed by the trade-off of intricate calculus for obscure 
geometry''.\cite{Tiberiis} There is, however, an alternative approach which requires only  
some elementary vector algebra and calculus for its justification.

\begin{figure*}[t!] 
\includegraphics[width=2.5in]{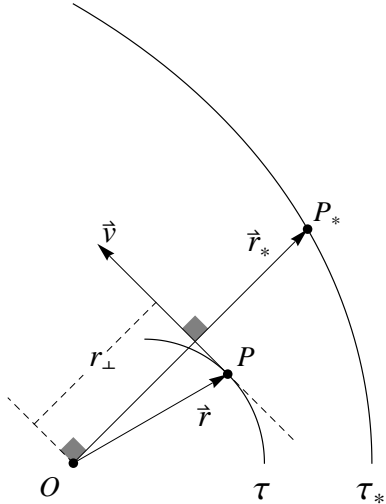}
\caption{The map $P\rightarrow P_*$ and the polar reciprocal $\tau_*$ of $\tau$; $r_*=r_\perp^{-1}$. 
(The notation $\mbox{\myorh{$r$}}_*$ for $\mbox{\myorh{$OP$}}_*$ is convenient in 
Appendix~\ref{app:involution}.)}\label{fig:defPR} 
\end{figure*}

The seed for this other construction can be traced back to a result of Newton 
(Proposition I, Corollary I on page 41 of Ref.~\onlinecite{Newton}) arising from the conservation of
angular momentum in a central force field. Let \overrightharp{$r$} be the position vector of a body relative to
the center $O$ of the force field and let \overrightharp{$C$} be the body's angular momentum per unit 
mass with respect to $O$; then, 
\begin{equation}\label{eq:C}
   C = |\mbox{\overrightharp{$C$}}| 
     = |\mbox{\overrightharp{$r$}}\times\dot{\mbox{\overrightharp{$r$}}}\,| 
    = r_\perp |\dot{\mbox{\overrightharp{$r$}}}\,|\, ,
\end{equation}
where $r_\perp$ is the component of \overrightharp{$r$} perpendicular to the instantaneous velocity
$\mbox{\overrightharp{$\nu $}}=\dot{\mbox{\overrightharp{$r$}}}$ (relative to $O$): since $C$ is a constant 
of the motion, the speed 
$\nu=|\dot{\mbox{\overrightharp{$r$}}}\,|$ of the body is inversely proportional to $r_\perp$ ($\nu=C/
r_\perp$). Consider now the mapping $P\rightarrow P_*$ illustrated in Fig.~\ref{fig:defPR}. 
The point $P$ is on a typical trajectory $\tau$ of a body in an (attractive) central force field (with 
center $O$); of course, $\tau$ is in the plane through $O$ perpendicular to \overrightharp{$C$} (and so are
\overrightharp{$r$} and \overrightharp{$\nu$}). By construction, the vector $\mbox{\overrightharp{$OP$}}_*$
(in this plane) is perpendicular to the velocity  \overrightharp{$\nu$} at $P$ and has magnitude
$|\mbox{\overrightharp{$OP$}}_*|=r_\perp^{-1}$. Since $r_\perp^{-1}=\nu/C$ [from Eq.~(\ref{eq:C})] and 
the direction of
$\mbox{\overrightharp{$OP$}}_*$ is obtained from that of $\mbox{\overrightharp{$\nu $}}$ by a clockwise 
rotation through $90^\circ$ (see Fig.~\ref{fig:defPR}), \emph{the curve $\tau_*$\/} in Fig.~\ref{fig:defPR},
which is the locus of points $P_*$ as $P$ varies over $\tau$, is \emph{the hodograph of $\tau$, rescaled 
by a factor of $C^{-1}$ and rotated clockwise through 90${}^\circ$\/} (see also section 3 of 
Ref.~\onlinecite{Chakerian}). Following Ref.~\onlinecite{Chakerian}, I will term $\tau_*$ the 
\emph{polar reciprocal\/} of $\tau$.

It is the inverse of the mapping $P\rightarrow P_*$ in Fig.~\ref{fig:defPR} which can be used to 
draw trajectories. In fact, there is a pleasing symmetry. The mapping $P\rightarrow P_*$ 
is involutory, i.e.\ $(P_*)_*=P\,$ (see Appendix~\ref{app:involution} for an elementary proof),
so trajectories in a central force field and their hodographs (after 
rotation and rescaling as in the preceding paragraph) are polar reciprocals of each other. \emph{The 
point-by-point determination of a trajectory from a rotated and rescaled hodograph involves exactly 
the steps depicted in Fig.~\ref{fig:defPR}}.\cite{Anosov}
Unlike the methods of Refs.~\onlinecite{Derbes} and \onlinecite{SalasBrito}, polar reciprocation is valid 
for \emph{any\/} smooth hodograph associated with \emph{any\/} central force field. Be that as it may, I 
will now specialize to orbits of the Kepler problem. I will also use the identity
\begin{equation}
 \mbox{\overrightharp{$OP$}}_*=\mbox{\overrightharp{$\nu $}} \times\mbox{\overrightharp{$C$}}/C^2 ,
\end{equation}
which takes advantage of the fact that, for the trajectory $\tau$ depicted in Fig.~\ref{fig:defPR}, 
\overrightharp{$C$} points perpendicularly out of the page, so that $\mbox{\overrightharp{$\nu $}} 
\times\mbox{\overrightharp{$C$}}$ is parallel to $\mbox{\overrightharp{$OP$}}_*$ and has magnitude $\nu\, C$.

\begin{figure*}[b!]
\includegraphics[width=3in]{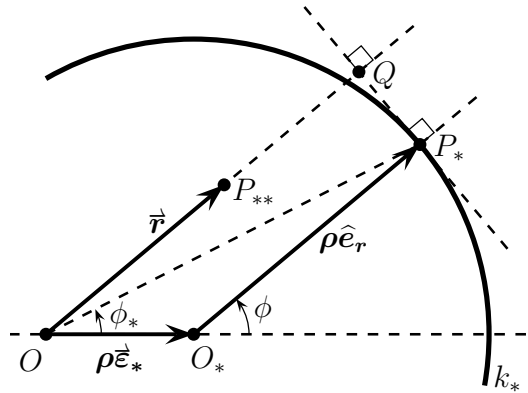}
\caption{A typical point $P_{**}$ on the polar reciprocal of $k_*$; $\angle P_*OP_{**}=\phi-\phi_*$ as
$O_*P_*\parallel OP_{**}$. By construction, $OP_{**}=OQ^{-1}=[OP_*\cos(\widehat{P_*OP_{**}})]^{-1}$ just as,
in Fig.~\ref{fig:defPR}, $OP_*=[OP\cos(\widehat{POP_*})]^{-1}$.}\label{fig:prK} 
\end{figure*}

For an inverse-square force per unit mass of $-\gamma \mbox{\overrightharp{$r$}}/r^3$ ($\gamma>0$), the
equation of motion for the position vector $\mbox{\overrightharp{$OP$}}_*$ of a typical point $P_*$ on
the polar reciprocal of an orbit (see Fig.~\ref{fig:defPR}) reads
\begin{equation}\label{eq:tdOPstar}
  \frac{d\ }{dt}\mbox{\overrightharp{$OP$}}_* 
    = \frac{d\mbox{\overrightharp{$\nu$}}}{dt} \times \frac{\mbox{\overrightharp{$C$}}}{C^2}
    = -\frac{\gamma}{Cr^2} \,\widehat{e}_r\times \frac{\mbox{\overrightharp{$C$}}}{C}
    = +\frac{\gamma}{C^2} \frac{d\phi}{dt}\, \widehat{e}_\phi
    = \rho \frac{d\widehat{e}_r}{dt} ,
\end{equation}
where $\rho\equiv\gamma/C^2$ and plane polar coordinates $r$ and $\phi$ have been adopted for the orbital 
plane; the corresponding unit vectors are $\widehat{e}_r$ and $\widehat{e}_\phi$ with the origin of the 
coordinate system at the force center $O$, and the azimuthal angle $\phi$ defined as in Fig.~\ref{fig:prK}
(so that $\mbox{\overrightharp{$C$}}=r^2\dot{\phi}\widehat{k}$).
Equation (\ref{eq:tdOPstar}) implies that $\mbox{\overrightharp{$OP$}}_*-\rho\widehat{e}_r$ is a (vectorial) 
constant of the motion, say $\mbox{\overrightharp{$OO$}}_*$ (drawn in Fig.~\ref{fig:prK}). Setting 
$\mbox{\overrightharp{$OO$}}_*= \rho {\mbox{\overrightharp{$\varepsilon $}}}_*$,
\begin{equation}\label{eq:Kstar} 
  \mbox{\overrightharp{$OP$}}_*  
     =  \rho (\mbox{\overrightharp{$\varepsilon$}}_* +\widehat{e}_r), 
\end{equation}
which demonstrates that the polar reciprocal $k_*$ of an orbit $k$ of the Kepler problem
is a circle or an arc of a circle (as suggested by the plot in Fig.~\ref{fig:prK}).\cite{Vivarelli}

The polar reciprocal of $k_*$ (which would be the corresponding orbit $k$) comprises points like $P_{**}$
in Fig.~\ref{fig:prK}, the image (under polar reciprocation) of $P_*$ on $k_*$. In terms of the angles in 
Fig.~\ref{fig:prK}, $r=OP_{**}=1/OQ=[OP_*\cos(\phi-\phi_*)]^{-1}=(\mbox{\overrightharp{$
OP$}}_*\cdot\widehat{e}_r)^{-1}$ or, evaluating $\mbox{\overrightharp{$OP$}}_*\cdot\widehat{e}_r$ 
with Eq.~(\ref{eq:Kstar}),
\begin{equation}\label{eq:rsqd}
   \frac{1}{r} = \rho (\mbox{\overrightharp{$\varepsilon$}}_*\cdot\widehat{e}_r + 1) 
               = \rho (1 + |\mbox{\overrightharp{$\varepsilon$}}_* |\cos\phi) .
\end{equation}
Comparison of Eq.~(\ref{eq:rsqd}) with the standard equation\cite{Coxeter} for a conic in polar 
coordinates confirms that $P_{**}$ is on a conic with 
focus $O$, eccentricity $e=|\mbox{\overrightharp{$ \varepsilon $}}_* |$, \emph{latus rectum\/}
$2/\rho$, and directrix perpendicular to $OO_*$. The angle $\phi$ can thus be identified as 
the \emph{true anomaly\/} (of celestial mechanics), and $\mbox{\overrightharp{$\varepsilon $}}_* $ can be
reinterpreted as a vector of magnitude equal to the eccentricity $e$ of the orbit $k$, directed 
from the force center $O$ to the point on $k$ of closest approach (i.e.\ the 
\emph{periapse}). In fact, to within a constant, $\mbox{\overrightharp{$\varepsilon $}}_* $ is the 
Laplace-Runge-Lenz vector.\cite{Goldstein}

In the previous two paragraphs, it has been shown that it is easy to establish the character of the polar
reciprocal $k_*$ of an orbit $k$ of the Kepler problem and even easier to infer from $k_*$ that the orbit $k$ 
must be a conic section. It is now possible to indicate how, as an alternative to the methods of 
Refs.~\onlinecite{Derbes} and \onlinecite{SalasBrito}, polar reciprocation may be used, in principle, 
to draw orbits with a compass and ruler.

Suppose that the orbiting body's velocity is given at some point $P_0$ 
on the orbit (which need not coincide with an apse of the orbit); let $\mbox{\overrightharp{$r$}}_0$ be the 
position vector of $P_0$ relative to $O$, and let the associated velocity (in the rest frame of $O$)
be $\mbox{\overrightharp{$\nu $}}_0$.  First, one must construct $k_*$. The following sequence of steps can
be employed (${P_0}_*$ below denotes the image under polar reciprocation of $P_0$):
\begin{itemize}
\item[(i)] calculate $C=|\mbox{\overrightharp{$r$}}_0\times\mbox{\overrightharp{$\nu $}}_0|$ and 
     then $\rho\,(=\gamma C^{-2})$; 
\item[(ii)] add graphically the vectors  ${\mbox{\overrightharp{$OP$}}_0}_* = 
     \mbox{\overrightharp{$\nu $}}_0\times\mbox{\overrightharp{$C $}}/C^2 $ and 
     $\mbox{\overrightharp{${P_0}_* O$}}_* =  - (\rho/r_0)\mbox{\overrightharp{$r$}}_0$ to find the position
     of $O_*$ relative to $O$; 
\item[(iii)] determine the eccentricity $e\,(=OO_*/\rho)$; 
\item[(iv)] at $O_*$, draw a circle of radius $\rho$ if $e<1$ or a circular arc on which the points $P_*$
     are such that $|\angle OO_*P_*| > \cos^{-1} e^{-1}$ if $e\ge 1$ (cf.\ Fig.~\ref{fig:PRHyperbola} in 
     Appendix \ref{app:suggestedProblems}).\cite{AppendixB} 
\end{itemize}
From a point $P_*$ on $k_*$, the corresponding point $P_{**}$ on $k$ can be determined by demanding that 
$\mbox{\overrightharp{$OP$}}_{**}$ is parallel to $\mbox{\overrightharp{$O_*P$}}_*$  and has magnitude
$|\mbox{\overrightharp{$OP$}}_{**}|=1/OQ$. Well-known compass-and-ruler constructions suffice. One can:
\begin{itemize}
\item[(1)] erect at $P_*$ a line $(l_{\perp})$ perpendicular to the line through $O_*P_*$;
\item[(2)] draw through $O$ the line $(l_{\|})$ parallel to the line segment $O_*P_*$, and then;
\item[(3)] invert the point of intersection of $l_{\perp}$ and $l_{\|}$ (the point $Q$ in Fig.~\ref{fig:prK}) 
     with respect to the unit circle centered on $O$.
\end{itemize}
Compass-and-ruler implementations (with \emph{JavaSketchpad}) of all three of these procedures
can be found at $<$\url{http://www.susqu.edu/brakke/constructions/constructions.htm}$>$ 
(accessed July 2, 2011).\cite{WallaceWest}

For students today, who have access to programs like \emph{Mathematica}, it may seem that there is little 
or no need for the geometrical construction under discussion; it takes only a few 
keystrokes (as in Example 2.1.9 of Ref.~\onlinecite{Tam}) to generate a plot of a Kepler problem orbit; 
I have, nevertheless, witnessed the intellectual satisfaction students experience in being able to 
correctly anticipate the character and orientation of an orbit: these features can be deduced from 
initial conditions via steps (i) to (iii) of the preceding paragraph. A more traditional alternative 
would be to use the Laplace-Runge-Lenz vector, but I have found that even my most talented and dedicated 
junior-level students are somewhat mystified by this construct and disinclined to use it, when it is 
introduced in isolation (as it usually is) as a single-valued combination of dynamical variables which happens
to be a constant of the motion. My students have been more receptive to this ``exotic'' constant of the 
motion when it is presented within the context of a discussion of hodographs or polar reciprocals of the 
Kepler problem. Other more mundane issues can be tackled. Some suggestions for problems are given in Appendix 
\ref{app:suggestedProblems}.

\begin{acknowledgments}
I would like to thank one of the anonymous referees of this paper for suggestions on improvements.
\end{acknowledgments}

\appendix

\section{PROOF THAT $\bm{P_{**}}\bm{=}\bm{P}$ IN A CENTRAL FORCE FIELD}\label{app:involution}

The proof below involves some elementary vector algebra and, crucially, use of the equations of motion, which 
can be assumed to be of the form
\begin{equation}\label{eq:EoM}
 \ddot{\mbox{\overrightharp{$r$}}} = f(r) \mbox{\overrightharp{$r$}} ,
\end{equation}
where $\mbox{\overrightharp{$r$}}(t) = x(t)\widehat{\imath}+y(t)\widehat{\jmath}$ is the position of the 
orbiting body relative to the force center (the orbital plane is taken the $xy$-plane with the $z$-axis 
parallel to the angular momentum per unit mass \mbox{\overrightharp{$C$}}).

By definition, under polar reciprocation, the point $P$ with position vector $\mbox{\overrightharp{$r$}}(t)$
(on trajectory $\tau$) is mapped to the point $P_*$ (on $\tau_*$) with position vector
$\mbox{\overrightharp{$r$}}_*(t)=\dot{\mbox{\overrightharp{$r$}}}(t) \times \mbox{\overrightharp{$C$}}/C^2$,
where $\mbox{\overrightharp{$C$}}=\mbox{\overrightharp{$r$}}(t)\times\dot{\mbox{\overrightharp{$r$}}}(t)$.
Likewise, under polar reciprocation, $P_*$ is mapped to the point $P_{**}$ with position vector
$\mbox{\overrightharp{$r$}}_{**} = \dot{\mbox{\overrightharp{$r$}}}_*(t)\times\mbox{\overrightharp{$C$}}_*/
C_*^2$, where $\mbox{\overrightharp{$C$}}_* = \mbox{\overrightharp{$r$}}_*(t)\times
\dot{\mbox{\overrightharp{$r$}}}_*(t)$. 

Substituting $\mbox{\overrightharp{$r$}}(t) = x(t)\widehat{\imath}
+y(t)\widehat{\jmath}$ into the definition of $\mbox{\overrightharp{$r$}}_*$, one finds that
\begin{equation}
 \mbox{\overrightharp{$r$}}_*(t) = \left[\dot{y}(t)\widehat{\imath}-\dot{x}(t)\widehat{\jmath}\,\right]/C ,
\end{equation}
where $C=x(t)\dot{y}(t)-y(t)\dot{x}(t)$ ($>0$). Using Eq.~(\ref{eq:EoM}) and the fact that $C$ is a constant 
of the motion, $\dot{\mbox{\overrightharp{$r$}}}_*$ can be simplified to
\begin{equation}\label{eq:rstar}
 \dot{\mbox{\overrightharp{$r$}}}_{*}(t) = f(r) \left[ y(t)\widehat{\imath}- x(t)\widehat{\jmath}\,\right]/C,
\end{equation}
and, hence,
\begin{equation}\label{eq:Cstar}
 \mbox{\overrightharp{$C$}}_* = -f(r)\widehat{k}/C.
\end{equation}
Equations~(\ref{eq:rstar}) and (\ref{eq:Cstar}) imply that $\dot{\mbox{\overrightharp{$r$}}}_{*}(t)
\times \mbox{\overrightharp{$C$}}_* = C_*^2\, \mbox{\overrightharp{$r$}}$ or 
$\mbox{\overrightharp{$r$}}_{**} = \mbox{\overrightharp{$r$}}$ (as required).

\section{SUGGESTED PROBLEMS}\label{app:suggestedProblems}

\noindent
(1) It is not unusual to see the orbital plane (or $xy$-plane) identified with the complex plane (or 
$z$-plane) via the 
(obvious) isomorphism $(x,y) \rightarrow z=x+iy$. Formulate polar reciprocation as an operation in the
complex plane. Use this representation to prove polar reciprocation is involutory.\cite{DEliseo}\\[1pc]
(2) The polar reciprocal $k_*$ of a known orbit $k$ of the \emph{Kepler problem\/} can be easily found. 
\hbox{a) Justify} the claim that the images under polar reciprocation of just two points on $k$ suffice to 
fix $k_*$.\\
b) Show that, with the $x$-axis aligned along $\mbox{\overrightharp{$\varepsilon $}}_*$, the velocity
at points around $k$ is
\begin{equation}\label{eq:vandphi}
 \mbox{\overrightharp{$\nu$}} 
           = (\gamma \rho)^\frac{1}{2} (e\, \widehat{\jmath} + \widehat{e}_\phi) 
\end{equation}
in the notation of this paper.\\
c) By considering the \emph{periapse\/} $A$ of $k$ (where $\mbox{\overrightharp{$\nu $}}_A\perp
\mbox{\overrightharp{$r$}}_A$ and $\mbox{\overrightharp{$r$}}_A  \parallel
\mbox{\overrightharp{$\varepsilon $}}_*$), demonstrate that the energy $E$ of the orbiting body is related 
to the eccentricity $e$ of $k$ by
\begin{equation}\label{eq:energy}
  \frac{E}{m} = {\textstyle\frac{1}{2}} \gamma\rho (e^2-1),
\end{equation}
where $m$ is the mass of the orbiting body. (Begin by finding expressions for 
$r_A^{-1}=\nu_A/C$ and  $\nu_A$ in terms of $\rho$ and $e$.)\\ 
d) Use Eqs.~(\ref{eq:vandphi}) and (\ref{eq:energy}) to prove that the 
inequality 
\(
\nu^2\ \left(=2E/m+2\gamma/r\right)> 2E/m
\)
reduces to $e\cos\phi >-1$.\\[1pc]
(3) The polar reciprocal of an elliptical orbit (of the Kepler problem) must be a \emph{complete\/} circle
(as opposed to only an arc of a circle) because of the periodicity of the motion. Show that 
this circle has radius 
\begin{equation}\label{eq:rhoE}
\rho = \frac{1}{2}\left(\frac{1}{r_p}+\frac{1}{r_a}\right), 
\end{equation}
where $r_p$ and $r_a$ are the distances (from the force center $O$) of the \emph{periapse\/} and 
\emph{apoapse}, respectively, and that the distance of its center $O_*$ from $O$ is
\begin{equation}\label{eq:OstarE}
OO_* = \frac{1}{2}\left(\frac{1}{r_p}-\frac{1}{r_a}\right) .
\end{equation}
Infer from Eqs.~(\ref{eq:rhoE}) and (\ref{eq:OstarE}) an expression for the eccentricity of the orbit.\\
\mbox{}

\begin{figure*}[t]
\includegraphics[height=2.25in]{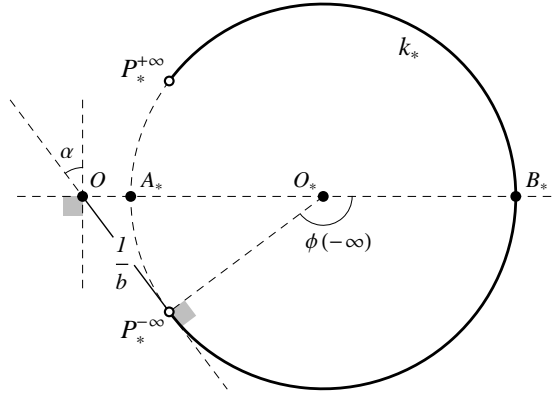}      
\caption{The polar reciprocal $k_*$ of the hyperbolic orbit $k$ of Fig.~\ref{fig:Hyperbola}. The point $B_*$
is the image under polar reciprocation of point $B$ in Fig.~\ref{fig:Hyperbola}; $P^{\pm\infty}_*$ are 
the images of the two \emph{points at infinity} $P^{\pm\infty}=\lim_{t\rightarrow\pm\infty}P(t)$ on the 
in/out asymptotes. Like $k$, $k_*$ is symmetric about the principal axis of $k$ (the line through $O$ and 
$O_*$).}\label{fig:PRHyperbola}
\end{figure*}

\begin{figure*}[t] 
\includegraphics[height=3.25in]{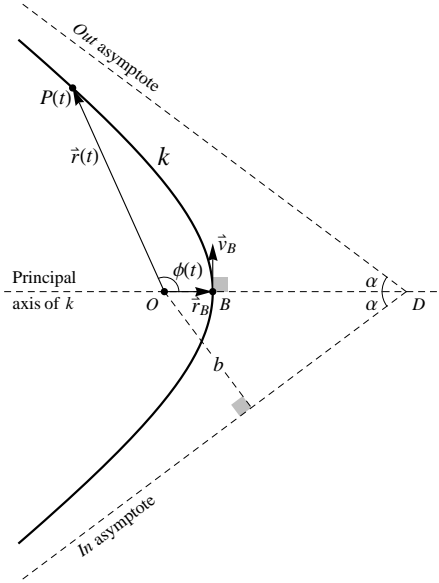}      
\caption{A hyperbolic orbit $k$ of the Kepler problem.}\label{fig:Hyperbola} 
\end{figure*}

\begin{figure*}[b]  
\includegraphics[width=2.25in]{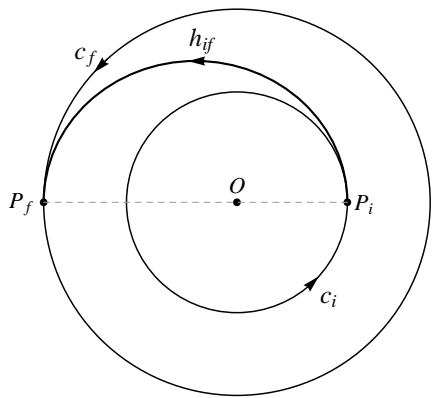}
\caption{The Hohmann transfer trajectory $h_{if}$ between coplanar circular orbits $c_i$ and $c_f$ in the 
field of a spherical mass centered on $O$; $h_{if}$ is tangent to $c_i$ at $P_i$ and tangent to $c_f$ at 
$P_f$.}\label{fig:HTO} 
\end{figure*}

\noindent
(4) Figure \ref{fig:PRHyperbola} depicts the polar reciprocal $k_*$ of the hyperbolic orbit $k$ in 
Fig.~\ref{fig:Hyperbola}. Prove that $OP_*^{-\infty}$ is tangent to the circle in 
Fig.~\ref{fig:PRHyperbola} of which $k_*$ is a part by showing that $\angle OP_*^{\pm\infty} O_*=90^\circ$.
By appealing to the ``tangent-secant theorem'',\cite{Euclid} deduce that $OA_* = r_B /b^2$.
Hence, conclude that, in Fig.~\ref{fig:PRHyperbola}, the circle has radius
\begin{equation}\label{eq:rhoH}
 \rho = \frac{1}{2r_B}\left(1 - \frac{r_B^2}{b^2}\right)
\end{equation}
and 
\begin{equation}\label{eq:OOstarH}
   OO_* = \frac{1}{2r_B}\left(1 + \frac{r_B^2}{b^2}\right) .
\end{equation}
Use Eqs.~(\ref{eq:rhoH}) and (\ref{eq:OOstarH}) to justify the assertion that results appropriate to 
parabolic orbits can be obtained by taking the limit $b\rightarrow\infty$, keeping $r_B$ fixed.
\mbox{}\\
(5) Many spacecraft maneuvers involve short firings of high-thrust rockets, which can be modeled as
resulting in instantaneous changes $\Delta\mbox{\overrightharp{$\nu $}}$ in velocity with no change in 
position. The important Hohmann transfer (depicted in Fig.~\ref{fig:HTO}) requires two such impulsive 
thrusts: the first (at $P_i$) places the vehicle on the semi-elliptical trajectory
$h_{if}$, and the second (at $P_f$) puts the vehicle into the circular orbit $c_f$.\cite{Gregory}
\mbox{}\\
a) Show that, provided $\Delta\mbox{\overrightharp{$\nu $}}$ is tangential to the spacecraft's orbit and
the direction of motion of the spacecraft is not reversed, 
$\mbox{\overrightharp{$r$}}_*=\mbox{\overrightharp{$\nu $}}
\times\mbox{\overrightharp{$C$}}/C^2$ is unchanged by an impulsive thrust.\\
b) Find the polar reciprocal representation of the Hohmann transfer in Fig.~\ref{fig:HTO}, i.e.\ draw
   (in one diagram) the polar reciprocals of $c_i$, $c_f$, and $h_{if}$. \\
c) Generalize the result of part b) to the case of two counter-clockwise non-intersecting co-planar and 
\emph{co-axial} elliptical orbits about the force center $O$. The transfer trajectory must connect the 
periapse of the inner orbit to the apoapse of the outer orbit.


\begin{thebibliography}{99}
\bibitem{Anosov} Trajectories for which $C=0$ have to be excluded, but these are one-dimensional and easily 
    drawn once the initial conditions are known. See D. V. Anosov,
    ``A note on the {K}epler problem," J. Dynamical and Control Systems \textbf{8}, 413-442 (2002).
\end{thebibliography}
\end{document}